# Enhanced Physicochemical and Biological Properties of Ion-Implanted Titanium Using Electron Cyclotron Resonance Ion Sources

Csaba Hegedűs [1], Chia-Che Ho [2], Attila Csik [3], Sándor Biri [3,], Shinn-Jyh Ding [2,4,]

1. Department of Biomaterials and Prosthetic Dentistry, University of Debrecen, Nagyerdei krt. 98, H-4032 Debrecen, Hungary; chegedus@edu.unideb.hu
2. Institute of Oral Science, Chung Shan Medical University, Taichung City 402, Taiwan; sfox1223@gmail.com
3. Institute for Nuclear Research (ATOMKI), Hungarian Academy of Sciences, Bem tér 18/c, H-4026 Debrecen, Hungary; csik.attila@atomki.mta.hu
4. Department of Dentistry, Chung Shan Medical University Hospital, Taichung City 402, Taiwan

**Abstract:** The surface properties of metallic implants play an important role in their clinical success. Improving upon the inherent shortcomings of Ti implants, such as poor bioactivity, is imperative for achieving clinical use. In this study, we have developed a Ti implant modified with Ca or dual Ca + Si ions on the surface using an electron cyclotron resonance ion source (ECRIS). The physicochemical and biological properties of ion-implanted Ti surfaces were analyzed using various analytical techniques, such as surface analyses, potentiodynamic polarization and cell culture. Experimental results indicated that a rough morphology was observed on the Ti substrate surface modified by ECRIS plasma ions. The *in vitro* electrochemical measurement results also indicated that the Ca + Si ion-implanted surface had a more beneficial and desired behavior than the pristine Ti substrate. Compared to the pristine Ti substrate, all ion-implanted samples had a lower hemolysis ratio. MG63 cells cultured on the high Ca and dual Ca + Si ion-implanted surfaces revealed significantly greater cell viability in comparison to the pristine Ti substrate. In conclusion, surface modification by electron cyclotron resonance Ca and Si ion sources could be an effective method for Ti implants.

## 1. Introduction

Titanium and titanium alloys are the metallic materials most commonly used in load-bearing applications because of their superior mechanical properties. Poor bioactivity is a significant shortcoming of metallic implants, which results in the failure of bonding directly to bone tissues. Many studies have been done for improving and optimizing metallic implant surfaces by creating bioactive coatings or modifying the surfaces, because the biomaterial surface plays a crucial role in biological interaction. To ensure successful medical use, the key physical properties of a metal implant material should be retained, while the outermost surface is modified to affect the interaction between the implant and its human host. Surface modification methods applied to the inert implants include texturing/passivation, anodization, ion implantation, silanization and deposition processing [1–7]. Surface texturing, such as sandblasting and acid etching, could increase the surface area of the implant, thereby enhancing osseointegration by increasing the area to which bone can bond [1]. In ion implantation, high-energy ions bombard the implant surface and enhance the corrosion resistance and/or biocompatibility of the titanium through the formation of a ceramic surface layer [2

The incorporation of osteoconductive or osteoinductive molecules on Ti surfaces is one of the effective modifications [7–11]. It has been documented that extracellular Ca is a potent regulator of cell behavior and has significant effects on the proliferation and differentiation of osteoblasts [12]. Sawada *et al.* found that Ca(OH)$_2$-treated titanium disks induced osteogenic differentiation in human mesenchymal stem cells [10]. Titanium implant surfaces modified with calcium ions using sonication dipping stimulate platelet adhesion and activation and provisional matrix formation [11]. Hanawa and co-workers found that Ca ion-implanted titanium in an amount of $10^{17}$ ions/cm$^2$ was superior to titanium alone for bone conduction after implantation in rat tibia [13]. In addition to Ca ion, Si species can enhance attachment, proliferation, differentiation and mineralization of osteoblast-like cells and mesenchymal stem cells [7,14], as well as osteogenesis *in vivo* [15].

It is not easy to find a suitable facility that is able to deliver calcium, silicon and other ion beams with the required charge, intensity and energy. During the process of finding the most optimal solution, finally, a special device is chosen, which has been used mainly in physical research and up-to-date has not spread in medical research. The electron cyclotron resonance (ECR) ion source (ECRIS) is used world-wide to deliver highly charged ions for particle accelerators to get very high energy beams at the end. Their many unique features make them one of the most promising devices to produce a wide variety of low energy ion beams for direct surface treatments. ECRIS can produce ion beams from many gaseous or solid materials. By changing the ion extraction voltage of the source and/or the charge state distribution of the plasma, the implantation depth of the ions into the materials can be varied easily. Usually, it should be between 1 and 100 nm, which is available applying standard terminal voltages. Thus, the surface properties of the materials only are modified, the bulk mechanical properties remaining unchanged. The Atomki-ECRIS [16,17] seems to be a suitable device for this task, because it is an independent ion source (there is no post-acceleration) offering beam time freely. Furthermore, the modularity of this source opens the possibility to change its configuration within a reasonable time. In this study, ECRIS consisting of Ca or Ca + Si ions was used to modify the Ti surfaces. Calcium and silicon ion vapors are ionized by electrons heated by microwave radiation. The microstructure, corrosion resistance and biocompatibility of ion-implanted Ti substrates were characterized.

## 2. Experimental Section

### 2.1. Preparation of Substrate

Grade 2 commercially available 1 mm-thick titanium plates (99.6 at%, Grade 2, Spemet Co., Taipei, Taiwan) of 10 × 10 mm$^2$ were selected as the substrate materials. Prior to ion implantation, the substrate surface was mechanically polished to #2000 grit level, followed by 1-µm Al$_2$O$_3$ powder to produce a mirror-like surface. The substrates were then etched in 30% HNO$_3$ for 30 min at room temperature, followed by ultrasonic cleaning in ethanol for 30 min, rinsing with distilled water for 30 min and air drying.

### 2.2. Ion Implantation

The Atomki-ECRIS (Room-temperature, 14.3 GHz; Atomki, Debrecen, Hungary) system can produce a variety of charged ion beams from H, He, CO, CH4, N, O, Ne, Ar, Kr and Xe gases and C, C$_{60}$, Fe, Ni, Zn and Au ions from solids. However, in the present study, new methods for calcium and silicon ion beam production were necessary to develop. A calcium plasma and ion beam was obtained using a commercial filament oven (PK10-0107-DT01, Pantechnik, Bayeux, France). The oven head was placed close to the plasma chamber of the ECRIS system and filled by pure calcium. Due to the vapor pressure of the calcium, the optimal operation temperature was found between 500 and 700 °C. Helium was used as the support gas. Multiply-charged calcium plasma was generated with some He background. A typical calcium ion beam spectrum is shown in Figure 1 [17]. The plasma was optimized Ca$^{3+}$ with the maximum extracted Faraday cup current (FCU) of about 20 eµA.

To obtain a silicon ion beam, we used SiH$_4$ (silane) gas. Special care had to be applied because of its highly flammable feature. A special gas handling system was designed to transfer the silane gas from a high volume and high pressure (2 dm$^3$ and 50 bar, respectively) gas bottle to a smaller (50 cm$^3$ and 1.5 bar, respectively) one. The small bottle was connected to the ion source through a gas dosing valve, and a silicon-hydrogen mixture ion beam was produced. Due to the safety arrangements, the risk was reduced to the normal operation level. For low and middle charge states ($Q$ = 1 to 8), silicon beam currents between 25 and 100 eµA were easily obtained (Figure 2), which gave us enough freedom to vary the kinetic energy of the beam and, thus, to regulate the implantation depth [17].

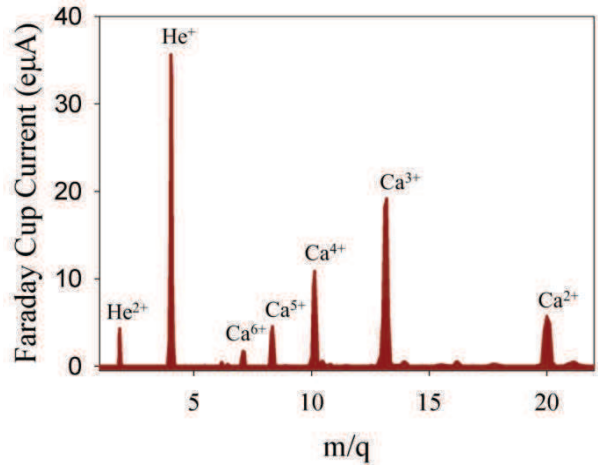

**Figure 1.** Calcium ion beam spectra. The extraction voltage and the temperature of the oven were 5 kV and 700 °C, respectively. The plasma was optimized for Ca$^{3+}$.

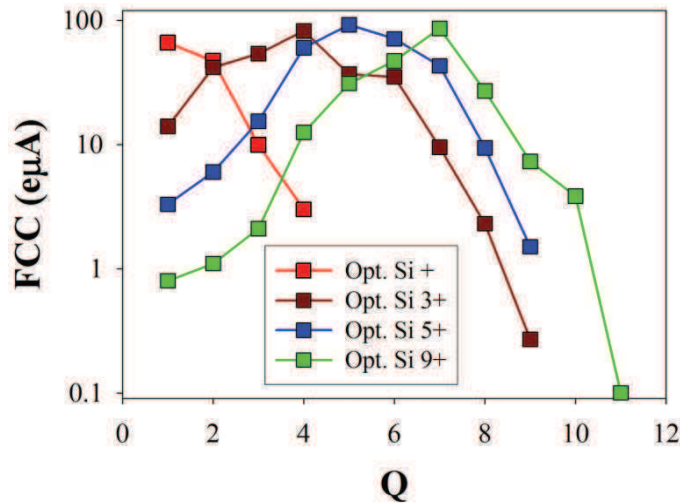

**Figure 2.** Charge state distributions of the analyzed silicon ion beam spectra as a function of the optimized charge states.

Besides the development of the necessary ion beams (Ca and Si), a new irradiation chamber was also designed and manufactured. Thus, the exchange of the materials to be irradiated became fast and easy without breaking the vacuum. The size of the ion beam was about 50 mm in diameter and circular. In order to reach as high Ca and Si dose as possible, the irradiation facility was built very close to the ECRIS. By this arrangement, all of the extracted components of the plasma reach and hit the samples, of course. We believe, however, that the effect of the unnecessary, but unavoidable components of the beam (mainly H and He) is negligible. The extraction voltage was 10 kV, and the ECRIS system was tuned for $Ca^{2+}$ or $Si^{2+}$ production. This meant that ions with charges from 1 up to 4 or 5 hit the samples, resulting in different implantation depths. Ti samples were implanted by $5 \times 10^{16}$ ion/cm$^2$ Ca, $1 \times 10^{17}$ ion/cm$^2$ Ca and $5 \times 10^{16}$ ion/cm$^2$ Ca + $5 \times 10^{16}$ ion/cm$^2$ Si dose, respectively, which were also denoted as 5Ca, 10Ca and 5CaSi, respectively.

*2.3. Composition and Morphology*

A secondary neutral mass spectrometry (SNMS) system (INA-X, SPECS GmbH, Berlin, Germany) was used for the depth profiling of the samples. Phase analysis of sample surfaces with and without ion implantation was performed using a grazing incidence X-ray diffractometer (GIXRD; Bruker D8 SSS, Bruker Corporation, Karlsruhe, Germany) with Ni-filtered CuKα radiation operating at 40 kV and 100 mA with a glancing incident angle at 0.5° and a scanning speed of 1 °/min. The chemical structure of the samples was analyzed using Fourier transform infrared spectroscopy (FTIR; Bruker Vertex 80v, Bruker Corporation) with a spectral resolution of 1 cm$^{-1}$ in reflection at grazing incidence θ = 80° and a wavenumber range of 400 to 4000 cm$^{-1}$. Surface morphologies of various samples were coated with gold and observed under a field emission scanning electron microscope (SEM; JEOL JSM-6700F, JEOL Ltd., Tokyo, Japan). The Surfcorder SE-40G profilometer (Kosaka Laboratory, Tokyo, Japan) was used to measure the surface roughness (Ra; arithmetic mean roughness) with a cutoff value of 0.25 mm, a measurement length of 1 mm and a drive speed of 0.1 mm/s. The Ra data provided for each group were the mean of five independent measurements.

*2.4. Corrosion Measurement*

The corrosion measurements included open circuit potential (OCP) time methods and potentiodynamic polarization in a non-deaerated simulated body fluid (SBF) solution, using a CHI 660A electrochemical system (CH Instrument, Austin, TX, USA). The SBF solution, the ionic composition of which is similar to that of human blood plasma, consisted of 7.9949 g NaCl, 0.3528 g NaHCO$_3$, 0.2235 g KCl, 0.147 g K$_2$HPO$_4$, 0.305 g MgCl$_2 \cdot$ 6H$_2$O, 0.2775 g CaCl$_2$ and 0.071 g Na$_2$SO$_4$ in 1000 mL distilled H$_2$O and was buffered to pH 7.4 with hydrochloric acid (HCl) and tris-hydroxymethyl aminomethane (Tris, CH$_2$OH)$_3$CNH$_2$) [18]. All chemicals used were of reagent grade and used as obtained. For OCP measurement, only two electrodes (working electrode and reference electrode) were involved, whereas for the potentiodynamic polarization method, a conventional three-electrode cell was used. A saturated calomel reference electrode (SCE) and a platinum counter electrode were employed. The sample surface was cleaned by distilled water. The evaluation of potentiodynamic polarization was started after immersion in SBF for 1 h. The scanned potential range varied from −1 to 1 V toward the anodic direction at a sweep rate of 1 mV/s in the Tafel mode. The current was recorded in the absence of stirring or gas bubbling into the electrolyte. The corrosion potential and corrosion current were provided after being analyzed by the software. The results were obtained from five separate experiments.

*2.5. Hemocompatibility*

A 5-mL whole-blood sample obtained from a healthy human donor was added to 10 mL of phosphate-buffered saline (PBS). The red blood cells (RBCs) were harvested by centrifugation of the diluted blood solution at 10,400 rpm for 10 min and were then washed with PBS 5 times. A total of $5 \times 10^7$ RBCs in 1 mL of PBS was incubated with different samples in 24-well plates for 6 h at 37 °C. Then, the solution was centrifuged to obtain the supernatant; the absorbance of the supernatant at 570 nm was measured by a Sunrise microtiter plate reader (Tecan Austria Gesellschaft, Salzburg, Austria) with a reference of 655 nm. The absorbance of RBCs in deionized water and PBS served as positive (Cp) and negative controls (Cn), respectively. The hemolysis ratio was calculated using the following equation: hemolysis (%) = (A − Cn)/(Cp − Cn) × 100%, where A represents the absorbance of the RBC solution incubated with the samples [19]. The presented data are representative of three independent experiments.

*2.6. Cytoskeleton Observation*

The implant biocompatibility was evaluated by incubating the samples with MG63 human osteoblast-like cells (BCRC 60279, Hsinchu, Taiwan). Before cell incubation, the samples were sterilized by soaking in a 75% ethanol solution and exposure to ultraviolet light for 2 h. MG63 cells were suspended in Dulbecco's Modified Eagle's Medium (DMEM; Gibco, Langley, OK, USA) containing 10% fetal bovine serum (FBS; GeneDireX, Las Vegas, NV, USA) and 1% penicillin/streptomycin solution (Caisson, North Logan, UT, USA). The MG63 suspensions ($2 \times 10^4$ cells per well) were directly seeded over each sample and placed in a 24-well plate. To clarify the effects of the various ion-implanted surfaces on cell attachment, the actin cytoskeleton was analyzed via fluorescence microscopy. After 6 h of incubation, the unbound cells were washed off with cold PBS, and the adherent cells were fixed in 4% p-formaldehyde (Sigma-Aldrich, St. Louis, MO, USA) for 30 min at room temperature and permeabilized with 0.1% Triton X-100 (Sigma-Aldrich) in PBS for 10 min. The cells were then incubated with phalloidin conjugated to Alexa Fluor 594 (Invitrogen, Grand Island, NY, USA) to stain the cytoskeleton for 1 h. Next, the nuclei was stained with 300 nM 4′,6-diamidino-2-phenylindole (DAPI) (Invitrogen) for 1 h. After washing three times with PBS, the cells were viewed under indirect fluorescence using a Zeiss Axioskop2 microscope (Carl Zeiss, Thornwood, NY, USA) at 200× magnification.

*2.7. Cell Viability*

Cell viability was performed with an initial cell density of 5000 cells/well on Days 1, 3 and 7. Fresh culture medium was replaced every 2 days. At the end of the culture time, the medium was removed, and the samples were washed twice with PBS. A total of 650 µL of 0.5 mg/mL MTT ((3-(4,5-dimethylthiazol-2-yl)-2,5-diphenyltetrazolium bromide; Sigma-Aldrich) in culture medium was added and incubated at 37 °C for 2.5 h. Subsequently, 500 µL of dimethylsulfoxide (Sigma-Aldrich) was added, and the absorbance at 570 nm was determined using a Sunrise Microtiter Reader. The data provided for each group were the mean of three independent samples.

*2.8. Statistical Analysis*

One-way analysis of variance (ANOVA) was used to evaluate significant differences between means in the measured data. In the event of a significant difference between test groups, this necessitated testing all of the possible differences via multiple comparisons that were characterized by considering any significant differences between all possible pairs of groups. Scheffé's multiple comparison testing was used to determine the significance of the standard deviations in the measured data from each specimen under different experimental conditions. In all cases, the results were considered statistically significant with a *p*-value of less than 0.05.

**3. Results and Discussion**

*3.1. Chemical Composition*

Figure 3A shows the GIXRD patterns of Ti surfaces before and after ion implantation. Similar GIXRD patterns were obtained for all of the samples. Three characteristic peaks located at around 35.1°, 38.4° and 40.1° can be attributed to the (100), (002) and (101) crystal faces of the Ti substrate phase [9]. Interestingly, the reduction of (100) reflection was found on the ion implanted surface. It seems that there were two broad peaks at around 35.9° and 42.4° visible in the two Ca-implanted samples (the 5Ca and 10Ca groups). This was possibly because of an ultrafine crystal size (amorphous phase) associated with the surface prepared by the implantation [5]. As also observed in Figure 3B, the FTIR profile of the ion-implanted samples was almost featureless, indicating a high degree of amorphousness. Krupa *et al.* also reported amorphous calcium-ion implanted Ti samples using a dose of $10^{17}$ Ca/cm$^2$ [20].

In Figure 4, SNMS shows the depth profiles of three elements (Ti, Ca and Si) as a function of depth. Surface contaminants, such as carbon, oxygen and nitrogen, were also identified during the SNMS analysis in both control and ion-implanted samples. However, for the sake of clarity, we did not show the depth distribution of these elements here. It is reasonable to consider that the concentration of Ti on the surface of all groups increased and reached the saturation slowly, due to the formation of a titanium-based mixture on the surface. Depth profiling of the samples showed the presence of Ca in the surface layer of all ion-implanted Ti, while the Si signal was appreciable on the 5CaSi surface (Figure 4C), as expected. The presence of Si was identified by SNMS, indicating a 25-nm average depth.

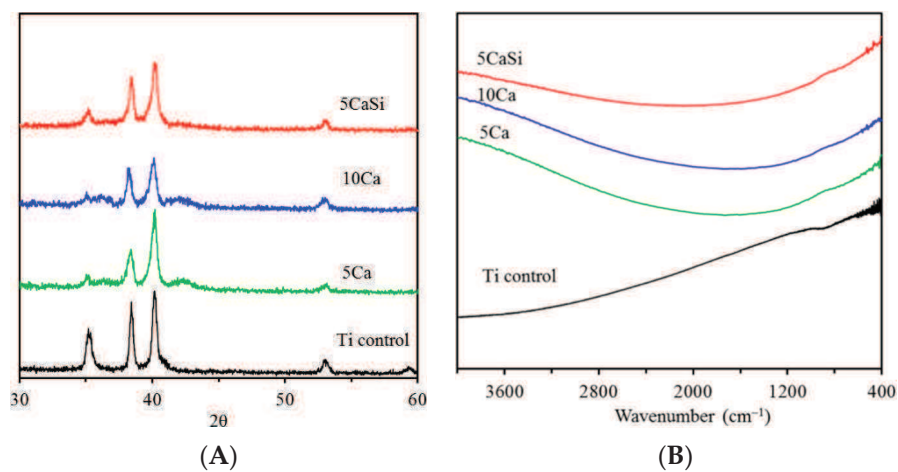

**Figure 3.** (**A**) Grazing incidence X-ray diffractometer (GIXRD) patterns and (**B**) FTIR spectra of the Ti samples with and without ion implantation.

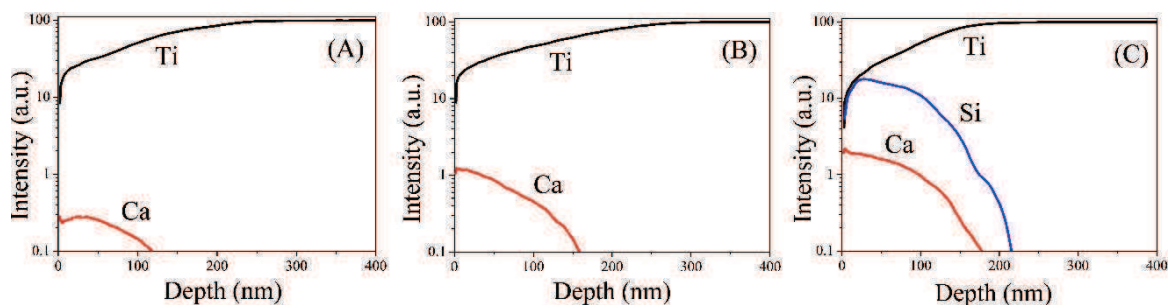

**Figure 4.** Secondary neutral mass spectrometry (SNMS) depth profiles of (**A**) 5Ca; (**B**) 10Ca and (**C**) 5CaSi ion-implanted surfaces.

*3.2. Morphology*

In general, the surface of Ti implants is pre-treated by sandblasting with a large grit and acid etching (SLA) for biomechanical interlocking that allows the bone ongrowth. However, the aim of the present study was to examine the ion implantation effect; relatively smooth sample surfaces were used throughout the study to reduce the interference of the roughened substrate and to facilitate the evaluation of the currently-used modification. The changed morphologies of the Ti substrate before and after ion implantation are shown in Figure 5. Compared to the pristine Ti substrate (Figure 5A), uniform distributed Ca granules with a size of about 300 nm built up the structure on the pristine Ti surface in a well-dispersed pattern (Figure 5B,C). In contrast, the incorporation of Si into the Ca-implanted surface induced numerous smaller granules over the Ti surface (Figure 5D), which could be distinguished from the Ca-implanted surfaces alone. Using this ion implantation, the smooth surface of the pristine Ti substrate was changed to a rougher morphology due to the formation of granules with a convex texture. As expected, the surface roughness results indicated a significantly ($p < 0.05$) higher Ra value ranging from 68 to 86 nm for ion-implanted groups compared to that obtained for the Ti control of 48 nm (Table 1). However, there were no significant differences ($p > 0.05$) in Ra roughness between the ion-implanted groups. The implantation process created surface changes that could result in altered topography and possibly even altered surface charge due to micro- or nano-scale pores or pits [21].

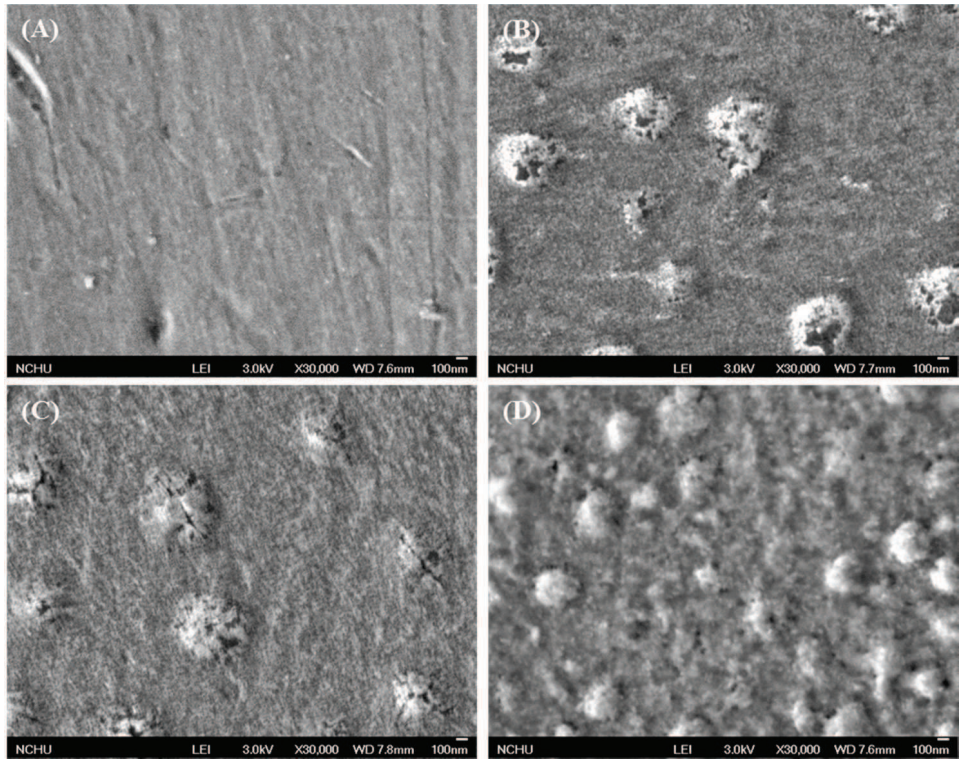

**Figure 5.** Surface SEM images of (**A**) Ti control; (**B**) 5Ca; (**C**) 10Ca and (**D**) 5CaSi groups.

**Table 1.** Roughness (Ra), electrochemical parameters and hemolysis ratio of all samples.

| Sample Code | Ra (nm) | Potentiodynamic Polarization | | Hemolysis (%) |
|---|---|---|---|---|
| | | $E_{corr}$ (mV) | $I_{corr}$ (nA) | |
| Control | 48 ± 8 [a] | −699 ± 114 [a] | 54 ± 27 [a,b] | 1.32 ± 0.16 [a] |
| 5Ca | 68 ± 8 [b] | −603 ± 98 [a,b] | 98 ± 22 [b,c] | 0.10 ± 0.23 [b] |
| 10Ca | 86 ± 11 [b] | −471 ± 84 [b] | 131 ± 55 [c] | 0.05 ± 0.38 [b] |
| 5CaSi | 76 ± 11 [b] | −575 ± 37 [a,b] | 30 ± 3 [a] | 0.16 ± 0.09 [b] |

Values are the mean ± standard deviation. Mean values followed by the same superscript letter (e.g., a, b, c) were not significantly different ($p > 0.05$) according to Scheffé's *post hoc* multiple comparison.

### 3.3. Corrosion Behavior

In order to evaluate the corrosion behavior of the Ti implants with and without ion modifications, the OCP over time of the samples in SBF is demonstrated in Figure 6. The OCP of all samples shifted towards the active direction after soaking in SBF, possibly due to the dissolution that occurred at the sample surfaces [4], followed by reaching a steady state. Compared to the Ti control, the ion-implanted samples showed a higher initial potential and attained more noble potential values, indicating a superior corrosion behavior. Among the samples, the 10Ca surface might exhibit a better corrosion resistance during the OCP measurements than the other sample surfaces.

Typical potentiodynamic polarization curves are presented in Figure 7. The measured corrosion potential ($E_{corr}$) and corrosion current ($I_{corr}$) were obtained using Tafel extrapolation methods, as summed in Table 1. For the corrosion potential, the $E_{corr}$ value of the pristine Ti substrate was −699 mV, while the $E_{corr}$ values of the ion-implanted samples ranged from −471 to −603 mV. Similar to the findings of OCP examination, the 10Ca group had a corrosion potential (−471 mV) significantly ($p < 0.05$) higher than that of Ti (−699 mV). However, no significant differences ($p > 0.05$) were observed among the three ion-implanted samples. Concerning the effect of Si on polarization, it seems not to affect the corrosion potential in comparison with the other two ion-implanted samples. Concerning $I_{corr}$, the comparative study displayed that the average values between 30 and 131 nA were dependent on the types of the samples. The 10Ca group had higher $I_{corr}$ values compared to the pristine Ti substrate and the 5CaSi group, possibly due to the dissolution of Ca ions when soaked in SBF. The released Ca ion may facilitate the cell growth. According to the literature [22], in tissue culture models, elevation in Ca concentration increases osteoblast chemotaxis and proliferation and alters the levels of expression of some differentiation markers. Compared to the pristine Ti substrate, the increased corrosion potential and the decreased corrosion current observed in 5CaSi samples can be explained by the structure of the surface layer being altered [20].

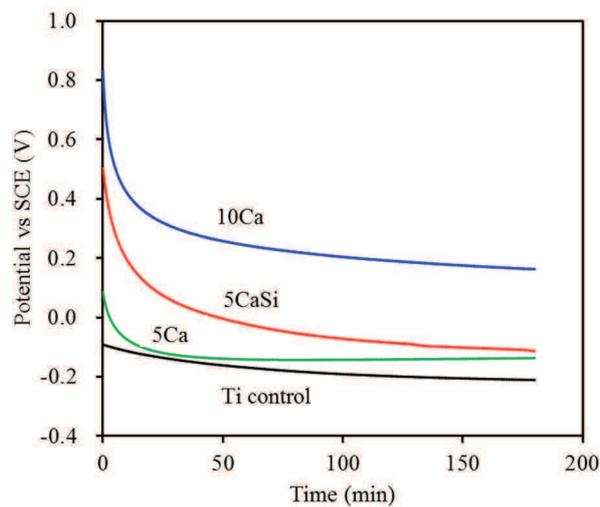

**Figure 6.** Open circuit potential (OCP) of Ti substrates with and without ion implantation. SCE, saturated calomel reference electrode.

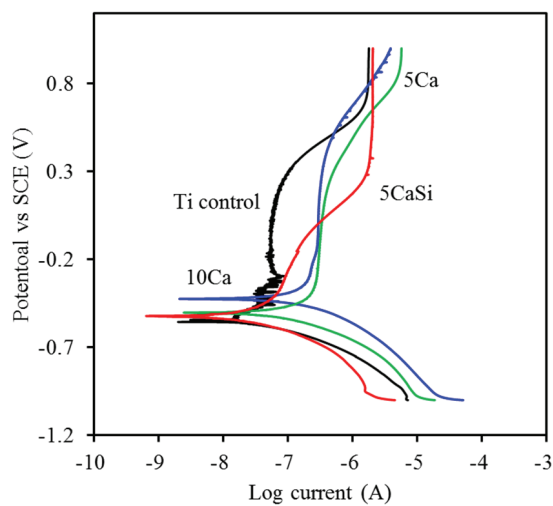

**Figure 7.** Typical potentiodynamic polarization curves of all samples in simulated body fluid (SBF).

*3.4. Hemocompatibility*

The hemolytic activities of all samples were evaluated by measuring the amount of hemoglobin released by ruptured RBCs. The results in Table 1 demonstrate that the hemolysis ratios of all samples are less than 2%, thus showing that these materials meet this criterion for biomedical materials (a hemolysis ratio lower than 5%) [19]. More importantly, all ion-implanted samples had a significantly lower ($p < 0.05$) hemolysis ratio than the Ti control. The hemolysis is linked to the surface characteristics of the test samples [19].

*3.5. Biocompatibility*

The biocompatibility of biomaterials is a prerequisite factor in the modulation of cellular function [23]. Surface chemistry is the most direct way to influence cell behavior [14]. To elucidate the effects of ion-implanted surfaces on biocompatibility, the responses of MG63 cells cultured to various samples were evaluated using fluorescent staining to actin. Actin stress fibers have been shown to act as a critical component contributing to cell migration [7]. The images revealed that after 6 h, MG63 cells cultured on all of the samples displayed highly tensioned actin stress fibers (red) (Figure 8). Cells were found growing in close proximity to all sample surfaces. Nayab *et al.* demonstrated that attachment and spreading of MG63 cells depended on the level of implanted Ca ions [24], which indicated a higher degree of cell spreading with increasing doses of Ca ions.

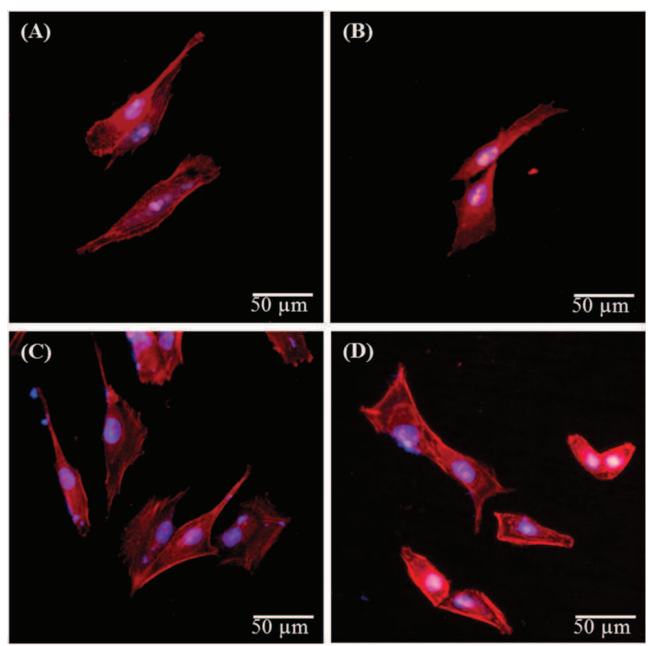

**Figure 8.** Fluorescent cytoskeleton staining of MG63 cells cultured on (**A**) Ti control; (**B**) 5Ca; (**C**) 10Ca and (**D**) 5CaSi after 6 h of seeding. The cells were stained for nuclei (blue) and actin cytoskeleton (red). Original magnification: 200×.

The quantification of adherent MG63 cells on different sample surfaces was performed using an MTT assay. The absorbance value steadily increased for all of the samples on Days 1 to 7, indicating increasing numbers of viable cells (Figure 9). There was no significant difference in the absorbance between all samples following culture for one and three days. However, cells rapidly attached to the 10Ca and 5CaSi surfaces after seven days compared to the pristine Ti and 5Ca groups. For example, the 5CaSi surface elicited a significant increment of 34% in comparison with the pristine Ti substrate after seven days of cell seeding. The 10Ca and 5CaSi implant surfaces could provide an optimal environment for osteoblasts to function normally. The cells seeded on the high Ca surface (10Ca group) for seven days had elevated metabolic activity, while cells incubated with a lower dose Ca surface (5Ca group) were not similarly activated. Krupa *et al.* also indicated that the viability and alkaline phosphatase bioactivity of the osteoblasts cultured in direct contact with non-implanted titanium and Ca-ion-implanted titanium (at a dose of $1 \times 10^{17}$ ion/cm$^2$) were the same [20]. Nevertheless, an appropriate Si concentration implanted onto the Ti surface was effective in supporting the proliferation of osteoblasts, in agreement with a recent study [7]. Last, but not least, the present results highlighted the beneficial effects of the incorporation of Si in the Ca surface layer on cell viability, which is worthy of further evaluation.

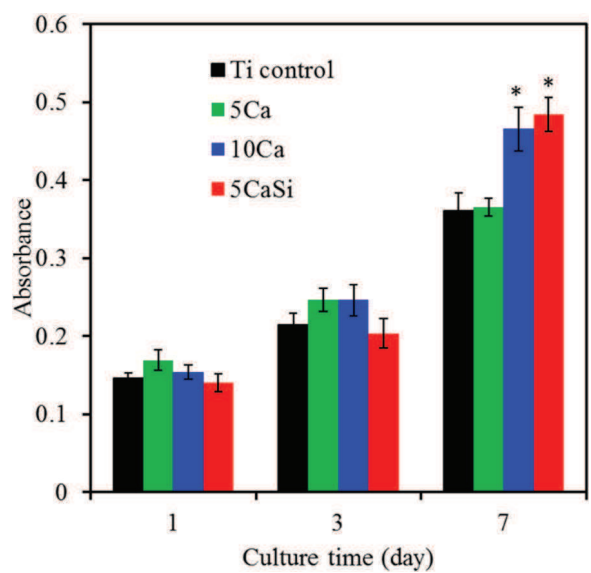

**Figure 9.** MTT assay for cell viability of MG63 cultured on the various samples after 1, 3 and 7 days of incubation. * Statistically significant difference ($p < 0.05$) from the Ti control.

## 4. Conclusions

The nature of the biomaterial surfaces may affect cell-material interactions, which in turn affect tissue development. Surface Ca and Ca/Si components were implanted onto the Ti metal by ECRIS. In light of the results obtained in this study, this kind of ion implantation technique did influence the surface topography of the pristine Ti substrate. The ion-implanted surface seemed to possess a better corrosion-resistant ability compared to the pristine Ti substrate. The improved surfaces did actively promote the viability of human osteoblast-like cells. The optimum protection on the Ti substrates was provided by a 10Ca or 5CaSi ion implantation in terms of the corrosion resistance and biocompatibility. Further studies, such as *in vitro* proliferation, differentiation and mineralization, are currently underway to evaluate the biological properties of the potential implant materials.

**Acknowledgments:** This work was partly supported by a Hungarian-Taiwan bilateral research program, No. NKM-92/2014, and MOST 103-2911-I-040-502. Work was also supported by the Hungarian TAMOP 4.2.2.A-11/1/KONV-2012-0036 project, which is co-financed by the European Union and the European Social Fund.

**Author Contributions:** Csaba Hegedűs was responsible for analyzing the results and writing the paper. Chia-Che Ho, Attila Csik and Sándor Biri performed the experiments. Shinn-Jyh Ding coordinated all tasks in the paper, planned the experiment and developed the results.

**Conflicts of Interest:** The authors declare no conflict of interest.